\documentclass[Phys,submission]{SciPost}

\usepackage[T1]{fontenc}
\usepackage{amsmath}
\usepackage{graphicx}
\usepackage{mathtools}
\usepackage{microtype}
\usepackage{xcolor}
\usepackage{float}
\hypersetup{hidelinks}

\newcommand{\onlinecite}[1]{\cite{#1}}

\def\thnc{{tHNC}}
\def\tVMC{{tVMC}}

\def\qf{{\bf f}}
\def\qv{{\bf v}}

\def\qr{{\bf r}}

\newif\ifcomment
\commentfalse

\newcommand{\JKUITP}{Institute for Theoretical Physics, Johannes Kepler University Linz,\\Altenberger Straße 69, 4040 Linz, Austria}
\newcommand{\FAU}{Institute for Quantum Gravity, Theoretical Physics III, Department of Physics, Friedrich-Alexander-Universität Erlangen-Nürnberg,\\Staudtstraße 7, 91058 Erlangen, Germany}
\newcommand{\JKUSSS}{Institute of Semiconductor and Solid State Physics, Johannes Kepler University Linz,\\Altenberger Straße 69, 4040 Linz, Austria}
\newcommand{\EPFL}{Institute of Physics, \'{E}cole Polytechnique F\'{e}d\'{e}rale de Lausanne (EPFL),\\CH-1015 Lausanne, Switzerland}

\begin{document}

\begin{center}{\Large \textbf{
Interaction quenches in Bose gases studied with a time-dependent hypernetted-chain Euler-Lagrange method
}}\end{center}

\begin{center}
Mathias Gartner\textsuperscript{1},
David Miesbauer\textsuperscript{1},
Michael Kobler\textsuperscript{2},
Julia Freund\textsuperscript{1,3},
Giuseppe Carleo\textsuperscript{4} and
Robert E. Zillich\textsuperscript{1*}
\end{center}

\begin{center}
{\bf 1} \JKUITP{}
\\
{\bf 2} \FAU{}
\\
{\bf 3} \JKUSSS{}
\\
{\bf 4} \EPFL{}
\\
\medskip
*\,robert.zillich@jku.at
\end{center}

\begin{center}
\today
\end{center}


\section*{Abstract}
{\bf 
We present a new variational method to study the dynamics of a closed bosonic many-body system, the time-dependent
hypernetted-chain Euler-Lagrange method, \thnc . Based on the Jastrow ansatz, it accounts for quantum fluctuations in a non-perturbative way. The \thnc\ method scales well with the number of dimensions, as demonstrated
by our results on one-, two-, and three-dimensional systems. We apply the \thnc\ method to interaction quenches, i.e.\ sudden
changes of the interaction strength, in homogeneous Bose gases. When the quench is strong enough that the final state
has roton excitations (as found and predicted for dipolar and Rydberg-dressed Bose-Einstein condensates, respectively),
the pair distribution function exhibits stable oscillations.
For validation, we compare \thnc\ results with time-dependent variational Monte Carlo results in one and two dimensions.
}

\vspace{10pt}
\noindent\rule{\textwidth}{1pt}
\tableofcontents\thispagestyle{fancy}
\noindent\rule{\textwidth}{1pt}
\vspace{10pt}


\section{Introduction}

The dynamics of many-body systems far from equilibrium and the role of interactions is
an interesting and intensely studied topic. Many phenomena not known from
linear response dynamics have been predicted and/or observed:
many-body localization, where interactions may prevent self-equilibration if a system with
disorder starts far from equilibrium~\cite{schreiber_observation_2015},
dynamical phase transitions~\cite{heyl_dynamical_2013} characterized by a non-analytical time evolution after a quench
near a quantum phase transition point~\cite{jurcevic_direct_2017,zhang_observation_2017},
orthogonality catastrophe of polarons after an interaction quench~\cite{mistakidis_quench_2019}, and
non-thermal fixed points predicted in relaxation dynamics~\cite{karl_strongly_2017,madeiraSymm22}.
In particular, the high level of control in experiments with ultracold quantum gases, either
as continuous gases in harmonic~\cite{dalfovoRMP99} and box traps\cite{navon_quantum_2021} or as lattice gases~\cite{morschRMP06},
facilitates the study of far from equilibrium dynamics~\cite{polkovnikovRMP11,langenAnnRevCondMat15},
by e.g. quenching an optical lattice~\cite{cheneau_light-cone-like_2012}
to study correlation dynamics, or by interaction quenches of Bose gases by Feshbach resonances~\cite{chinRMP10} to study
three-body correlations~\cite{fletcher_two-_2017} and universality~\cite{makotynNatPhys14}.
In the latter cases, a rapid quench to large $s$-wave scattering length $a$ is essential to investigate
strongly interacting Bose-Einstein condensates (BEC), because equilibrium studies are hampered by
losses due to three-body collisions when $\rho a^3$ becomes large, where $\rho$ is the number density.

Theoretical studies of many-body dynamics far from equilibrium have been performed by a variety of methods,
some of which are best suited for lattice systems, like time-evolving block decimation~(TEBD)~\cite{vidal2004prl},
the application of the time-dependent variational principle~(TDVP)~\cite{kerman1976aop} to matrix product states~(MPS)~\cite{haegeman2011prl},
non-equilibrium dynamical mean-field theory~(DMFT)~\cite{aoki2014rmp}, and the time-dependent density
matrix renormalization group method~(tDMRG)~\cite{white1992prl,feiguin2005prb,schollwock2005rmp}.
Some of them work best in one dimension, like methods based on continuous matrix product
states~(cMPS)~\cite{verstraete2010prl}, while others scale well to two and three dimensions,
such as multiconfigurational time-dependent Hartree approaches~(\mbox{MCTDHF})~\cite{lode2020rmp}.

The hypernetted-chain Euler-Lagrange (HNC-EL) method has been formulated for finding optimized ground states~\cite{Kro86,KroTrieste,QMBT00Polls} and the dynamics in the linear response regime, 
the latter also termed correlated basis function method~\cite{campbell2015prb}.
In this work, we derive an efficient time-dependent variational method for continuous Bose systems in
any dimension by generalizing the HNC-EL method to a fully time-dependent method. This time-dependent
hypernetted-chain Euler-Lagrange (\thnc ) method is based on a Jastrow ansatz for the wave function like
the ground state HNC-EL method,
and akin to the time-dependent variational Monte Carlo (tVMC) method~\cite{carleo2012sr,carleo2014pra,carleo2017prx}.
The \thnc\ method, however, can be orders of magnitude more efficient computationally because it does not
require Monte Carlo sampling.

In this work, we use \thnc{} to study the dynamics of a homogeneous Bose gas
after a sudden interaction quench.
We are interested in the short period of time after the quench where three-body losses are not
dominant yet, therefore we can neglect these losses and have a closed quantum system.
Our primary interest is the time evolution of the pair distribution function $g(\qr,t)$ after
the interaction quench, in particular after a quench to a strongly correlated system
exhibiting roton excitations, which have zero group velocity.
To assess the validity of the approximations of \thnc{} 
we compare our results in one and two dimensions with tVMC simulations.

\section{The time-dependent Hypernetted-Chain Euler-Lagrange method}

We consider a Bose gas of particles with positions $\qr_i$ and mass $m$ in $d$ dimensions,
interacting via a pair potential $v$. In equilibrium, such a system is
described by a time-independent Hamiltonian
\begin{align}
    H_0 = -\frac{\hbar^2}{2m} \sum_{j = 1}^N \Delta_j + \frac{1}{2} \sum_{k \neq l}^N v(\qr_k-\qr_l).
    \label{eq:H0}
\end{align}
In this work, we consider completely homogeneous systems, which for quantum gases can be approximately
realized by box traps~\cite{navon_quantum_2021}. We do not model the box trap boundaries where
the density falls off rapidly; instead, we consider only the constant-density part in the interior of the trap.
Consequently, the Hamiltonian~$H_0$ for our model system does not contain an external one-body potential.

For Bose symmetry, the ground state wave function
$\Phi_0(\qr_1,\dots,\qr_N)$ can be readily calculated using, for example,
exact quantum Monte Carlo simulations.
Variational approximations to $\Phi_0$ can be obtained with less computational
effort, such as variational Monte Carlo, including recent advances using artificial neural networks~\cite{dawid2022}.

A well-established and straightforward variational treatment that includes correlations,
i.e. ``quantum fluctuations'', in a non-perturbative way, are based
on the Jastrow-Feenberg ansatz and its generalizations. The many-body Bose ground state
$\Phi_0$ can be expressed in terms of two-body, three-body, etc. correlations,
\begin{align*}
  \Phi_0(\qr_1,\dots,\qr_N) = & \frac{1}{\sqrt{\mathcal{N}}}\, \exp\Big[\frac{1}{2}\sum_{k<l} u_2^{(0)}(\qr_k-\qr_l)
  +
  \frac{1}{3!}\sum_{k<l<m} u_3^{(0)}(\qr_k,\qr_l,\qr_m) + \dots \Big],
\end{align*}
where $\mathcal{N}$ denotes the normalization integral $\langle\Phi_0 | \Phi_0\rangle$.
The real-valued correlation functions~$u_n^{(0)}$ are obtained from the Ritz variational
principle, which requires that the energy expectation value
$E=\langle\Phi_0|H|\Phi_0\rangle/\langle\Phi_0|\Phi_0\rangle$ is minimized.
The series of correlations has to be truncated for practical calculations.
The Euler-Lagrange equations resulting from functional optimization, ${\delta E\over\delta u_n^{(0)}}=0$,
involve high-dimensional integrals, which can be evaluated approximately using
diagrammatic methods~\cite{HaM76}.
These equations are the hypernetted-chain Euler-Lagrange (HNC-EL) equations (the notation becomes clear below).
If two-body and three-body correlations are considered, the ground state energy and structural properties of strongly correlated systems, such as
liquid $^4$He, are very close to exact Monte Carlo results~\cite{Kro86}.
For a less strongly correlated system, two-body correlations are sufficient.

Excitations of the many-particle system described by $H_0$ can be obtained
from linear response theory, by allowing for small time-dependent fluctuations
of the correlations,
$u_n(\qr_k,\qr_l,\dots,t)=u_n^{(0)}(\qr_k,\qr_l,\dots)+\delta u_n(\qr_k,\qr_l,\dots,t)$,
and expanding the Euler-Lagrange equations up to linear order in $\delta u_n$,
see Ref.~\onlinecite{QMBT00Saarela} for details.
Typically, excitations are generated by probing the system with a weak external
time-dependent one-body potential, $\sum_j v_{\rm ext}(\qr_j,t)$ (such as a laser, neutrons etc.).
Then {one-body} ``correlations'' $\delta u_1(\qr_j,t)$ need to be included as well.  Excellent agreement
with experiments can be achieved, for example, for the dispersion relations of collective
excitations in $^4$He if fluctuations of three-body correlations are taken into
account~\cite{campbell2015prb}.

An interesting question is what happens if the external perturbation
is not weak, such that $\delta u_n(\qr_k,\qr_l,\dots,t)$ cannot be assumed to be
a small fluctuation around the ground state $u_n^{(0)}(\qr_k,\qr_l,\dots)$ anymore.
There are many examples of nonlinear response, such as
nonadiabatic alignment of molecules~\cite{stapelfeldt_colloquium:_2003,chatterley_rotational_2020},
dynamic material design~\cite{fechnerPRMat18,tulzer_quantum_2020},
or rapid parametric changes in ultracold gases such as interaction quenches~\cite{fletcher_two-_2017,villa_quench_2021},
the latter being the focus of the present work.
The nonlinear response of the system could be captured by expanding
the Euler-Lagrange equations to higher orders in $\delta u_n$.  Instead of
following this path, we want to formulate the Euler-Lagrange equations for
general time-dependent, complex correlations $u_n(\qr_k,\qr_l,\dots,t)$,
in order to find the time-dependent many-body wave function $\Phi(\qr_1,\dots,\qr_N,t)$
as an approximate solution of the time-dependent Schrödinger equation
$H(t)\Phi=i\hbar\dot\Phi$.
In this work, we focus on perturbations caused by changing the interaction potential,
which is modeled with the time-dependent Hamiltonian
\begin{equation}
  H(t) = -\frac{\hbar^2}{2m} \sum_{j = 1}^N \Delta_j + {1\over 2} \sum_{k \neq l}^N v(\qr_k-\qr_l,t).
\label{eq:Ht}
\end{equation}
No assumption about the magnitude of variations of $v$ in time
will be made.  Since we focus here on time-dependent interactions instead of time-dependent
external perturbation potentials, we restrict ourselves to homogeneous systems.
The interaction is translationally invariant, therefore the system
remains homogeneous despite the variation of $v$.  Regarding experimental realizations,
Feshbach resonances are one of the means to vary the effective interaction
over many orders of magnitude in experiments with ultracold quantum gases.

As long as the system is not too strongly correlated, two-body correlations are usually sufficiently
accurate.  Therefore, we restrict ourselves to two-body correlations $u_2(\qr_k-\qr_l,t)$ to discuss
the dynamics resulting from a quench of $v(\qr_k-\qr_l,t)$. It turns out to be convenient
to split $u_2$ into its real and imaginary part, $u_2(r)\equiv u(r)+2i\varphi(r)$,
where $r=|\qr_k-\qr_l|$ is the distance between particle $k$ and $l$.
The time-dependent generalization of the Bose Jastrow-Feenberg ansatz is
\begin{align}
    \Phi(\qr_1,\dots,\qr_N,t) =
  {1\over \sqrt{\mathcal{N}(t)}}\,\exp\Big[&{1\over 2}\sum_{k<l} u(|\qr_k-\qr_l|,t)
  + \, i\sum_{k<l} \varphi(|\qr_k-\qr_l|,t)\Big].
  \label{eq:Phi}
\end{align}
Since $u$ and $\varphi$ depend on time, all quantities introduced below depend on time as well.

The Euler-Lagrange equations of motion for $u$ and $\varphi$ are obtained from the generalization
of the Ritz variational principle to the time-dependent Schrödinger equation,
the minimization of the action integral
\begin{equation}
  {\mathcal{S}=\int_{t_0}^{t}\!dt' \mathcal{L}(t')}
\label{eq:defS}
\end{equation}
with the Lagrangian
\begin{equation}
	\mathcal{L}(t) = \langle\Phi(t) | H(t)- i \hbar\frac{\partial}{\partial t}|\Phi(t)\rangle.
	\label{eq:defL}
\end{equation}
The second expression in $\mathcal{S}$ involving the time derivative can be simplified
by the invariance of $\mathcal{S}$ with respect to adding total time derivatives to $\mathcal{L}$.
Terms involving the kinetic energy operator can be simplified with the Jackson-Feenberg identity for
a real-valued $F$
\begin{align*}
	F \Delta F &= \frac{1}{2}\big(\Delta F^2 + F^2 \Delta \big) +
	\frac{1}{2} F^2 \lbrack \nabla, \lbrack \nabla, \ln F \rbrack \rbrack
	-\frac{1}{4} \lbrack \nabla , \lbrack \nabla, F^2 \rbrack \rbrack,
\end{align*}
where $F=e^{{1\over 2}\sum_{k<l} u(|\qr_k-\qr_l|)}$ in our case.
The Lagrangian $\mathcal{L}$ can be brought into a convenient form, 
\begin{align}
  \mathcal{L} &=
  \mathcal{L}_g  + \mathcal{L}_3
  + {\hbar\over 2} \int\!\! d^dr\, \dot{\varphi}(r,t)\, g(r,t) 
  +{\hbar^2\over 2m} \rho\! \int\!\! d^dr\, \qv(\qr,t)^2 g(r,t),
   \label{eq:lagrangian}
\end{align}
with
\begin{align*}
  \mathcal{L}_g &=  {\rho\over 2} \int\!\! d^dr \Big[ v(r,t)
  - {\hbar^2\over 4m} \Delta u(r,t) \Big] g(r,t), \\
  \mathcal{L}_3 &=
  {\hbar^2\over 2m} \rho^2\! \int\!\! d^drd^dr'\, \qv(\qr,t) \cdot\qv(\qr',t)\,  g_3(r,r',|\qr-\qr'|,t).
\end{align*}
We abbreviated the gradient of the phase as ${\qv(\qr,t)\equiv \nabla\varphi(r,t)}$. 
The pair and three-body distribution functions, $g(r,t)$ and $g_3(r,r',|\qr-\qr'|,t)$,
are expressed as function of distance vectors (where $\qr=\qr_1-\qr_2$ and $\qr'=\qr_1-\qr_3$ are distances
between the particles at $\qr_1$, $\qr_2$, and $\qr_3$).
They are obtained from the corresponding pair and three-body densities via
\begin{align*}
  \rho_2(r,t) &= \rho^2 g(r,t),\\
  \rho_3(r,r',|\qr-\qr'|,t)& =\rho^3 g_3(r,r',|\qr-\qr'|,t),
\end{align*}
where the $n$-body density $\rho_n$ is defined as
\begin{align*}
\rho_n( \qr_1, \dots, \qr_n,t) = \frac{N!}{(N-n)!} \int\! dr_{n+1}\dots dr_N |\Phi(t)|^2.
\end{align*}
For a homogeneous system, $\rho_1\equiv\rho$ is the (constant) number density.

We vary $\mathcal{S}$ by functional derivation, for which we need the
relation between $u$ on the one hand and $g$ and $g_3$ on the other hand.
Since $g$ (and $g_3$) does not depend on $\varphi$, $g$ and $u$ are related via the
hypernetted-chain equation~\cite{HaM76} just as in ground state HNC-EL calculations,
\begin{align}
    g(r,t)=e^{u(r,t)+N(r,t)+E(r,t)},
\label{eq:HNC}
\end{align}
where $N(r,t)$ are the so-called nodal diagrams and $E(r,t)$ the elementary diagrams.  The former
can in turn be expressed in terms of $g$ via the Ornstein-Zernicke relation,
while the latter have to be approximated by truncating the infinite
series of elementary diagrams.  Details on these diagrammatic summations can be found in Ref.~\onlinecite{HaM76,Kro86}.
Note that the contribution $\mathcal{L}_g$, which does not depend on $\varphi$,
is just the expression for the energy expectation value also used in the ground state optimization
of the HNC-EL method.

Unlike the ground state energy expectation value for the Jastrow-Feenberg ansatz with pair correlations,
the Lagrangian $\mathcal{L}$ for the time-dependent
problem contains the three-body distribution $g_3$, which is a functional of $u$ and hence of $g$,
but cannot be given in a closed form. 
Two approximations of $g_3$ are common~\cite{feenberg1969}: the convolution
approximation which reproduces the correct long-range behavior of $g_3$, and the Kirkwood
superposition approximation
\begin{align}
  g_3(r,r',|\qr-\qr'|,t)\approx g(r,t)\,g(r',t)\,g(|\qr-\qr'|,t),
  \label{eq:kirkwood}
\end{align}
which reproduces the correct short-range behavior. Both
approximations can be systematically improved.  Since we
are interested here in impulsive changes of the short-range repulsion of the interaction,
we choose the latter approximation.

The time-evolution of the time-dependent Jastrow-Feenberg ansatz
$\Phi(t)$, eq.~(\ref{eq:Phi}), is determined by solving the time-dependent hypernetted-chain Euler-Lagrange
(\thnc ) equations
\begin{align}
  {\delta\mathcal{S}\over \delta g(r,t)} &= 0 \label{eq:deltaSdeltag}, \\
  {\delta\mathcal{S}\over \delta \varphi(r,t)} &= 0 \label{eq:deltaSdeltaphi}.
\end{align}
The \thnc\ equations are nonlinear partial differential equations. They could
be cast into a form which resembles the Navier-Stokes equations,
similar to the Madelung formulation of the one-body Schrödinger equation,
but we opt instead for a formulation in terms of a suitably defined ``wave function''
that is numerically more convenient.

For numerical solution of eqns.~(\ref{eq:deltaSdeltag}) and~(\ref{eq:deltaSdeltaphi}), we
define an effective ``pair wave function''
\begin{align*}
  \psi(r,t)\equiv \sqrt{g(r,t)}e^{i\varphi(r,t)}.
\end{align*}
Within the Kirkwood superposition approximation we can then cast the two real-valued 
eqns.~(\ref{eq:deltaSdeltag}) and~(\ref{eq:deltaSdeltaphi}) into a single complex nonlinear Schrödinger-like 
equation for $\psi(r,t)$.
The derivation can be found in the appendix. The final form of the \thnc\ equation is
\begin{align}
  i\hbar{\partial\over\partial t} \psi(r,t)
  =
  &- e^{-i\gamma(r,t)}{\hbar^2\over m}\nabla^2 e^{i\gamma(r,t)} \psi(r,t) \nonumber\\
  &+ [v(r,t)+w_I(r,t)]\, \psi(r,t) \nonumber\\
  &+ \Big[\beta(r,t) - {\hbar^2\over m}|\nabla \gamma(r,t)|^2\Big]\psi(r,t) .
\label{eq:tHNCel}
\end{align}
The induced interaction $w_I$ is a functional of $g(r)$ only, and it is the same expression that appears also in
ground state HNC-EL calculations, see e.g.\ Ref.~\onlinecite{QMBT00Polls}.
This induced interaction can be interpreted as a phonon-mediated interaction in addition to the bare interaction $v$. An additional
potential term appears if the elementary diagrams $E$ mentioned above are
taken into account; since $E$ depends only on $g$, the
same approximations for $E$ as in the ground state HNC-EL method could be applied. For
simplicity we neglect $E$ in this work.
The expressions $\beta$ and $\gamma$ are functionals of both $g(r)$ and $\varphi$,
\begin{align}
  \beta(r,t) &= {\hbar^2\over m}\rho\!\int\! d^dr'\ \qf(\qr',t) \cdot \qf(\qr'-\qr,t) \label{eq:beta},\\
  \nabla\gamma(r,t) &= \rho\!\int\! d^dr'\ \qf(\qr',t)\, [g(\qr'-\qr,t)-1] \label{eq:gamma},
\end{align}
with $\qf(\qr,t)\equiv g(r,t)\,\qv(\qr,t)$.  
Note that $\gamma$ in eq.~(\ref{eq:tHNCel}) is calculated by integrating $\nabla\gamma(r,t)$ given
in eq.~(\ref{eq:gamma}).

The formulation~(\ref{eq:tHNCel}) is chosen because it can be solved with standard
techniques such as operator splitting methods~\cite{chinPRE05}. 
In the zero-density limit $\rho\to 0$, all many-body effects vanish:
$\gamma\to 0$, $w_I\to 0$, $\beta\to 0$, hence eq.~(\ref{eq:tHNCel}) becomes the bare
two-body scattering equation for a two-body wave function $\psi$ with reduced mass ${m\over 2}$ at vanishing
energy. We remind that, in this paper, we have an isotropic, homogeneous interaction $v(|\qr_1-\qr_2|,t)$ and we
assume that the system remains isotropic and homogeneous for all times, i.e.\ it never
spontaneously breaks translation symmetry.
Lifting the restriction of homogeneity and isotropy is formally straightforward, but solving the
resulting \thnc\ equations would be computationally much more demanding.


\section{Results}

We present results for a homogeneous $d$-dimensional gas of bosons, where $d=1$, $2$, and $3$.
The interactions $v(r,t)$ are either simple models for a repulsive interaction or interactions between
Rydberg-dressed atoms~\cite{pohl_dynamical_2010}. In all cases, $v(r,t)$ is
characterized by two parameters: an interaction range $R$ and an interaction strength $U$,
see below.  The system is in the ground state $\Psi_0$ for times $t<0$, with interaction
parameters $R_0$ and $U_0$.  At $t=0$, we switch either the width parameter $R$
to a new value, $R(t)=R_0+(R_1-R_0)\Theta(t)$, or make a similar switch of $U$.
For $t>0$, the previous ground state $\Psi_0$ evolves according to the new
Hamiltonian $H$ characterized by the new interaction. At $t=0$ the energy changes
abruptly, but for $t>0$ the evolution is unitary and thus energy is conserved, because $H$ is
time-independent after the quench. Since there is no external potential and the interaction is
translationally invariant, the system is homogeneous before the quench; precluding symmetry breaking,
the system stays homogeneous after the quench.

In section~\ref{ssec:3D} we present \thnc\ results of the pair distribution function $g(r,t)$ after a quench in a Rydberg-dressed Bose gas
in three dimensions, where we show that roton excitations play an important role for the time-evolution
of the pair distribution function. Rydberg-dressed Bose gases
have been studied theoretically quite extensively~\cite{cinti_supersolid_2010,henkel_three-dimensional_2010,seydi_rotons_2020},
including studies of the dynamics after interaction quenches in the Bogoliubov approximation~\cite{mccormack_dynamical_2020},
and calculations of the roton excitation spectrum~\cite{henkel_three-dimensional_2010,seydi_rotons_2020}.
The influence of roton excitations on the dynamics after interaction quenches have been
studied for dipolar gases~\cite{natu_dynamics_2014}, again in Bogoliubov approximation.
In appendix~\ref{sec:bogo}, we compare the ground state $g(r)$ for the 3D Rydberg gas obtained within the Bogoliubov
approximation to results obtained with HNC-EL and with exact path integral Monte Carlo (PIMC) simulations in the
low temperature limit from~\cite{seydi_rotons_2020}. The comparison shows that HNC-EL agrees very well with
the exact PIMC result, while the Bogoliubov approximation
deviates quite strongly from the exact result.

In order to assess the approximations of the \thnc\ method, we compare the dynamics of $g(r,t)$
with tVMC results in two and one dimensions in section~\ref{ssec:compare}, using the same Jastrow ansatz
as in \thnc , but with fewer approximations. In general, correlations play a larger role in lower dimensional
systems, hence these comparisons are a harder test of \thnc\ than in three dimensions, and furthermore,
3D \tVMC\ simulations would have been computationally even more expensive.

\begin{figure}[ht]
\begin{center}
\includegraphics*[width=0.49\textwidth]{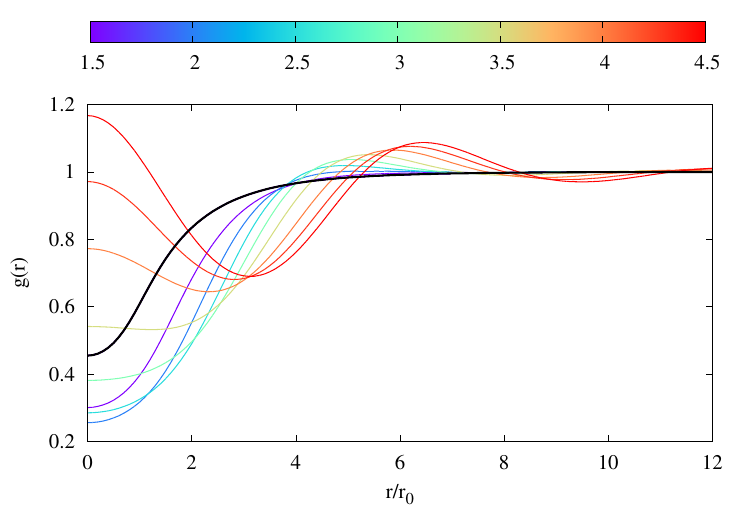}
\includegraphics*[width=0.49\textwidth]{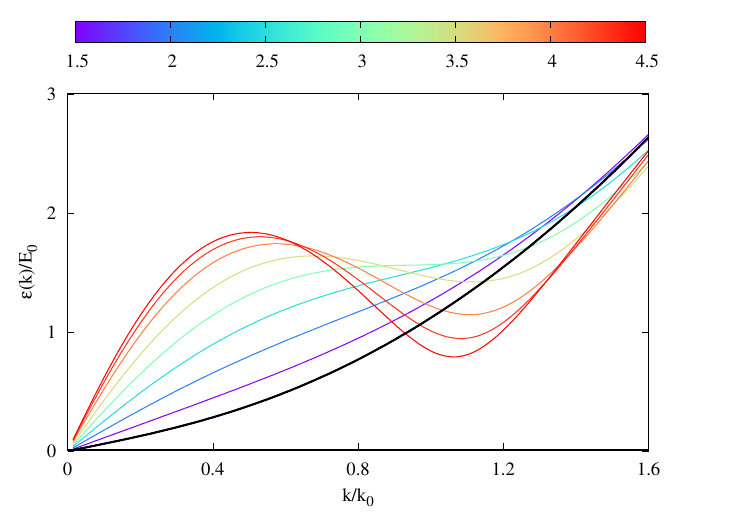}
\end{center}
\caption{
Left:
pair distribution function $g(r)$ for the ground state of a three-dimensional Rydberg-dressed Bose gas for several range parameters $R$
and for fixed interaction strength $U/E_0=2$, see eq.~(\ref{eq:vRyd}).
The thick black curve is for $R/r_0=1$, which is the initial value from which we quench to larger $R_1$ in
the dynamical calculations. Other curves are for $R/r_0=1.5;2.0;2.5;3.0;3.5;4.0;4.3;4.5$ (the colors correspond
to different $R$, given by the colorscale), which correspond to the target values $R_1$ to which we quench in the
dynamical calculations. Right:
the Bijl-Feynman excitation spectrum $\varepsilon_F(k)$ for the above values of $R$.
}
\label{FIG:3D199eps}
\end{figure}

\subsection{Quench dynamics in 3D}
\label{ssec:3D}

Rydberg-dressing means that the ground state and a Rydberg state of the atoms
are coupled by a laser detuned from resonance. The coupled
potential energy surfaces of the ground states and the Rydberg states
can be described by the following pair interaction
\cite{pohl_dynamical_2010,pupillo2010prl,cinti_supersolid_2010}
\begin{align}
    v(r,t) &= {U(t)\over 1+(r/R(t))^6}
    \label{eq:vRyd}
\end{align}
where the strength $U(t)$ and the range $R(t)$ may depend on time. When two particles
are closer than~${\approx R}$, they are only weakly repelled because $v$ becomes
flat for small $r$; for $r \gtrsim R$, they feel a van der Waals repulsion.
We follow Ref.~\onlinecite{seydi_rotons_2020} and measure wave numbers in units of $k_0=(6\pi^2\rho)^{1/3}$
in 3D, corresponding to a length unit $r_0=1/k_0$
(i.e.\ the density is always $\rho r_0^3=(6\pi^2)^{-1}$). Energy is measured in units
of $E_0={\hbar^2k_0^2\over 2m}$, and time in units $t_0={\hbar\over E_0}$.

\begin{figure}[ht]
\begin{center}
\includegraphics*[width=1\textwidth]{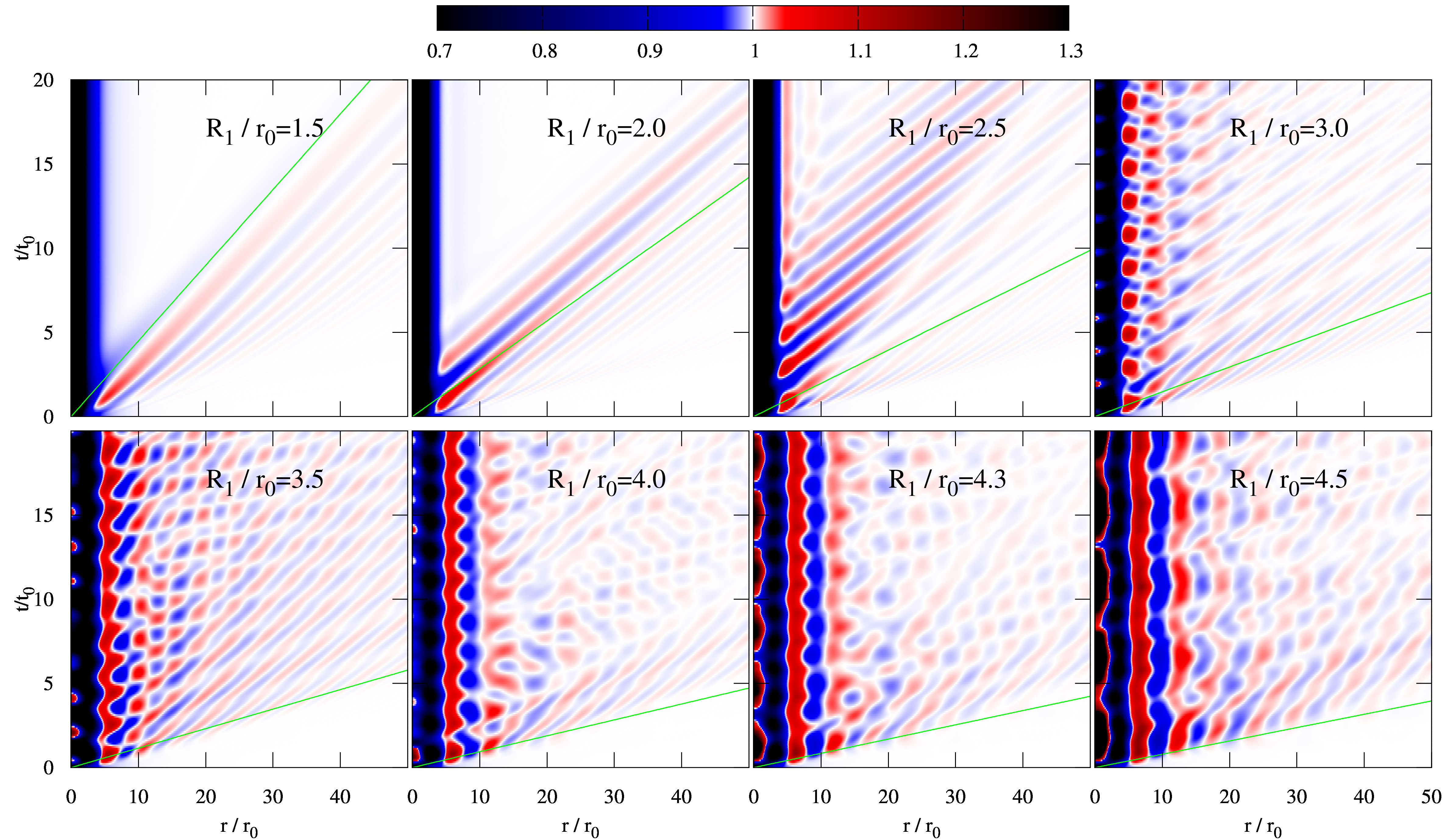}
\end{center}
\caption{
Pair distribution function $g(r,t)$ of a 3D Rydberg-dressed Bose gas after a quench from
$R_0/r_0=1$ to $R_1/r_0=1.5;2.0;2.5;3.0;3.5;4.0;4.3;4.5$, with fixed $U/E_0=2$. Blue indicates $g(r,t)<1$
and red indicates $g(r,t)\ge 1$ (see color scale). For $R_1/r_0\approx 3.0$ and higher, the oscillations of $g(r,t)$ near $r=0$
do not decay anymore due to generation of roton pairs with vanishing group velocity. The green line shows the
``sound cone'' $r=2ct$, where $c$ is the speed of sound after the quench.
}
\label{FIG:3D399}
\end{figure}

We study quenches from weak to strong interactions.
We keep $U$ fixed at $U_0=2\,E_0$, but switch $R(t)$ at $t=0$, from $R_0/r_0=1$ to $R_1/r_0=1.5;2.0;2.5;3.0;3.5;4.0;4.3;4.5$.
In the left panel of Fig.~\ref{FIG:3D199eps} we show the corresponding {\em ground state} pair density distributions $g(r)$.
The spatial oscillations in $g(r)$ become more pronounced as $R_1$ grows, and the range of correlations
increases. Typically for the interaction~(\ref{eq:vRyd}), particles tend to cluster for larger $R$,
eventually leading to a cluster solid~\cite{cinti_supersolid_2010}. This tendency to cluster
is seen in the growth of $g(r)$ for small distance $r$ as we increase $R_1$.

The time evolution of $g(r,t)$ after a quench from ${R_0/r_0=1}$ to the target values is shown
in Fig.~\ref{FIG:3D399} as color maps, where the horizontal axis is the distance $r$ and the vertical
axis is the time $t$. The evolution of $g(r,t)$ shows that the information about the
quench of the pair interaction is spreading to larger distances $r$ as time evolves.
Lieb and Robinson proposed a ``light cone'' outside of which the effect of the quench has
not yet arrived~\cite{lieb_finite_1972}. This light cone bound applies only to discrete Hamiltonians, and
was found for the Bose Hubbard model in Ref.~\onlinecite{carleo2014pra} with tVMC in one and two dimensions,
where the ``light'' are the elementary excitations.
Such a light cone appears if the group velocity $v_g$ of the elementary excitations has an upper
bound $c$. Then a quench can excite two waves in opposite directions (conserving total momentum zero);
the information about the quench would travel with $2c$, hence the response of $g(r,t)$ is expected to move with $2c$ to larger $r$.

The green lines in Fig.~\ref{FIG:3D399} show the light cone based on the speed of sound $c$ of the
Rydberg-dressed Bose gas, the ``sound cone'' given by $r=2ct$.
Especially up to $R_1=3\,r_0$ the waves in
$g(r,t)$ move faster than $2c$, and are not bound by the sound cone. This is not surprising since
$c$ is not the highest group velocity. Especially for low $R_1$, this can be seen
in the right panel of Fig.~\ref{FIG:3D199eps}.
There we show the excitation spectrum $\varepsilon_F(k)$ for the initial $R_0$ before
the quench (thick line) and for the target values $R_1$ (colored lines),
using the Bijl-Feynman approximation~\cite{bijl1940p,feynman1956pr}, $\varepsilon_F(k)={\hbar^2 k^2\over 2m\,S(k)}$.
The Bijl-Feynman spectrum is calculated with the ground state static structure factor $S(k)$ obtained
from a ground state HNC-EL/0 calculation with the respective value $R_1$.
For larger $R_1$, the dispersion relation $\varepsilon_F(k)$ becomes steep, corresponding to a large $c$.
In this regime, the evolutions of $g(r,t)$ approximately obey the sound cone bound, but upon closer inspection one
can see small oscillations with large wave number which spread faster than $2ct$. Hence, the sound cone is not a strict bound.
Again, this is not surprising because, at least in the Bijl-Feynman approximation, the dispersion relation
of the Rydberg-dressed Bose gas has arbitrarily large
group velocities for large $k$, because $\varepsilon_F(k)\to{\hbar^2 k^2\over 2m}$ for $k\to\infty$.
Note that this is different from excitations in lattice Hamiltonians, characterized by quasi-momenta within the
finite Brillouin zone, and thus $v_g$ does have a maximum (in the usual single-band
Hubbard approximation). The group velocity $v_g$ is usually the sound velocity, which becomes the maximal speed of
information spreading.

\begin{figure}[ht]
\begin{center}
\includegraphics*[width=0.68\textwidth]{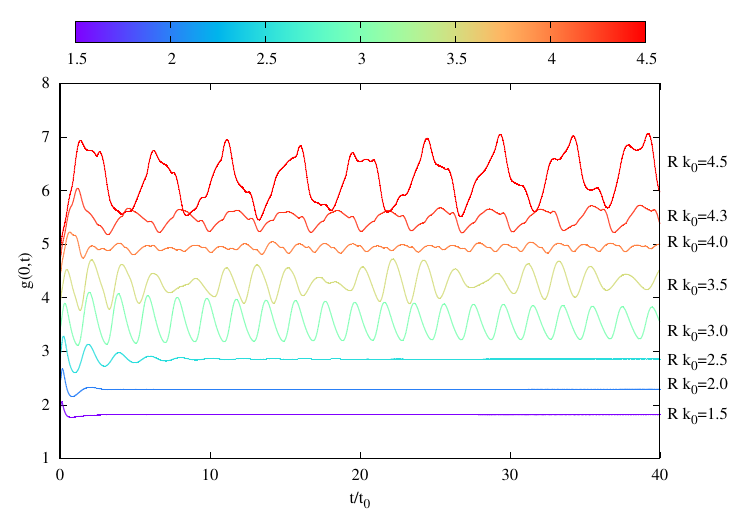}
\end{center}
\caption{
The pair distribution at $r=0$, $g(0,t)$, after a quench from $R_0/r_0=1$ to $R_1/r_0=1.5;2.0;2.5;3.0;3.5;4.0;4.3;4.5$, with
fixed $U/E_0=2$. For better visibility, the curves are shifted with respect to each other and colored based on $R_1$, see
colorscale at the top of the figure. For $R_1/r_0=3$ and higher,
$g(0,t)$ oscillates with no apparent decay, while for smaller $R_1$, the pair distribution at zero distance 
equilibrates to a constant value.
}
\label{FIG:3Dr0}
\end{figure}

For small target values $R_1$, $g(r,t)$ quickly converges to an equilibrium distribution in an $r$-interval
that grows with time, as the perturbation travels away to large $r$. The equilibrium is not
the ground state $g(r)$ at the same $R_1$, because the quench injects energy
into the system. For example for the quench from $R_0=1\,r_0$ to $R_1=2\,r_0$, the energy per particle jumps from
$E=0.073\,E_0$ (the ground state energy before the quench) to $E=0.739\,E_0$ and then of course stays constant
during the unitary time evolution; the {\em ground state} energy for $R_1=2\,r_0$, however, is lower, $E_g=0.643\,E_0$.

When we quench the interaction to larger $R_1$, there is a qualitative
change in $g(r,t)$: we observe long-lived oscillations
for $R_1 \ge 3\,r_0$ that do not decay within the time window shown in Fig.~\ref{FIG:3D399}.  This can be seen, for example, for small $r$. In Fig.~\ref{FIG:3Dr0} we show $g(r,t)$ for $r=0$ for all target values $R_1$.
The pair distribution function at $r=0$ is one way to obtain the contact
parameter~\cite{tan_energetics_2008,werner_general_2012-1} which can be measured~\cite{wild_measurements_2012}.
For $R_1 \le 2.5\,r_0$, $g(0,t)$ converges to a constant value, which lies slightly above the ground state $g(0)$.
For $R_1\ge 3\,r_0$, $g(0,t)$ appears to keep oscillating indefinitely.
Apart from $R_1=3\,r_0$, these oscillations clearly contain more than one frequency.

The origin of this long-lived oscillation for small $r$ becomes apparent
when we invoke the picture of a quench that generates two opposite excitations.
For small $R_1$, the Bijl-Feynman spectrum $\varepsilon_F(k)$ increases monotonically with wave number $k$,
see right panel of Fig.~\ref{FIG:3D199eps}.  For $R_1=3\,r_0$, $\varepsilon_F(k)$
has a plateau with essentially zero slope around $k/k_0=1$, and for larger $R_1$,
$\varepsilon_F(k)$ exhibits a maximum, called maxon, with energy $\hbar\omega_m$, and a minimum, called roton,
with energy $\hbar\omega_r$. A vanishing group velocity
$v_g(k)={d\varepsilon_F(k)\over dk}$ implies a diverging density of state, leading to a high probability
to excite excitations with $v_g\approx 0$. Furthermore, the excitation pairs of opposite momenta produced by the quench
do not propagate for $v_g=0$.  For $R_1>3\,r_0$, roton pairs as well as maxon pairs with opposite momenta are generated.
Since they do not propagate, the temporal oscillations for small distance $r$ become long-lived for $R_1\ge 3\,r_0$.

\begin{figure}[ht]
\begin{center}
\includegraphics*[width=0.68\textwidth]{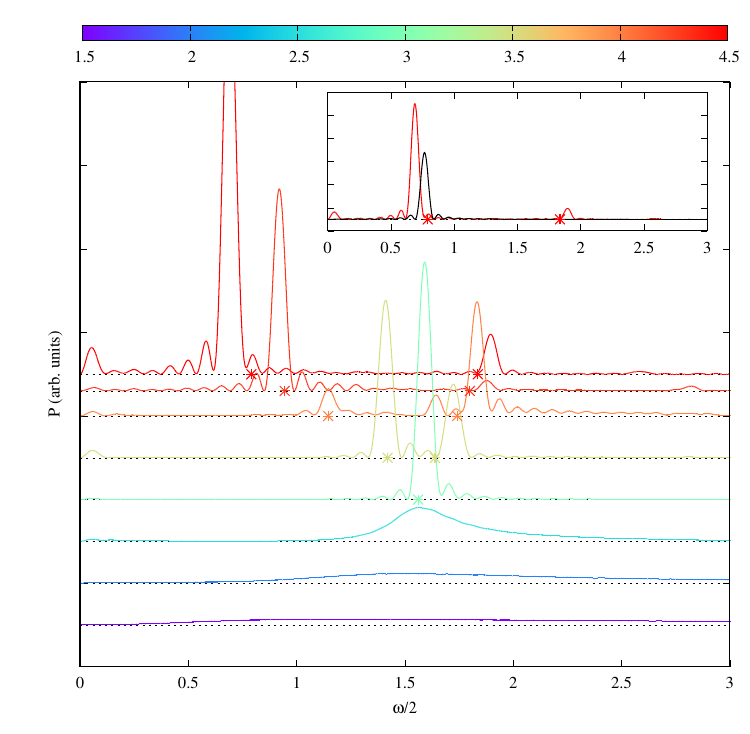}
\end{center}
\caption{
Power spectrum $P(\omega)$ of $g(0,t)$ shown in Fig.~\ref{FIG:3Dr0}, after a quench from $R_0/r_0=1$ to
$R_1/r_0=1.5;2.0;2.5;3.0;3.5;4.0;4.3;4.5$, with fixed $U/E_0=2$. For better visibility,
the curves are shifted with respect to each other and colored based on $R_1$, see
colorscale at the top of the figure.
The stars on the base line indicate the energies of the roton and the maxon after the quench, if present.
The inset compares $P(\omega)$ after a quench to $R_1=4.5\,r_0$ starting from $R_0=r_0$ (red) with a quench
starting from $R_0=4\,r_0$ (black).
}
\label{FIG:g0power}
\end{figure}

For $R_1=3\,r_0$, $g(0,t)$ oscillates with a single frequency because maxon and roton coincide
at the inflection point of $\varepsilon_F(k)$. The frequency is twice the corresponding excitation energy
(because the quench produces a {\em pair} of excitations).  For $R_1>3\,r_0$, $g(0,t)$ oscillates with
two frequencies, given by twice the roton and twice the maxon frequency.  In order to confirm this
quantitatively, we show the power spectra $P(\omega)$ of $g(0,t)$ in Fig.~\ref{FIG:g0power},
as functions of $\omega/2$. Each $P(\omega)$ is shifted in proportion to $R_1$ for better visibility.
For $R_1<3\,r_0$, the power spectra are broad. At $R_1=3\,r_0$ a single peak appears, corresponding to
the single frequency oscillations seen in Fig.~\ref{FIG:3Dr0}. For $R_1>3\,r_0$, $P(\omega)$ has two peaks of
varying relative spectral weight: at twice the roton frequency $2\,\omega_r$ and at twice the maxon
frequency $2\,\omega_m$ (the combination $\omega_r+\omega_m$ would have finite momentum and cannot
be excited in this simple picture by a translationally invariant perturbation such as an interaction quench).
The small ringing oscillations are artifacts from the Fourier transformation of a finite time window $[0,40\,t_0]$.

However, $\omega_r$ and $\omega_m$, which are indicated by stars in Fig.~\ref{FIG:g0power} for the respective $R_1$,
do not match perfectly with the peaks of $P(\omega)$. The picture of an interaction quench exciting two elementary excitations
with opposite momenta is only approximately valid. A quench from $R_0/r_0=1$ to e.g. $R_1/r_0=4.5$ is a highly
nonlinear process that cannot be regarded as a small perturbation and treated with linear response theory. The quench
tends to shift the lower-frequency below $2\,\omega_r$ and the higher-frequency peak
above $2\,\omega_m$. The effect of nonlinearity is demonstrated in the inset, where the power spectrum
$P(\omega)$ for the quench~$1\,r_0\to 4.5\,r_0$ shown in the main
figure (red) is compared with $P(\omega)$ for a much weaker quench~$4\,r_0\to 4.5\,r_0$ (black), where linear response theory may hold. From linear response theory, we would expect that, if two rotons are created, $g(r,t)$ will oscillate with exactly twice the roton frequency. This is indeed what we observe in the inset:
the power spectrum has a peak vert close to $2\,\omega_r$. Note that for the weaker quench, the excitation of two maxons is completely
suppressed, because the quench injects less energy into the system. The difference between the energy/particle after the weak quench
and the energy/particle of the $R_1/r_0=4.5$ ground state is just $\Delta E=0.03\,E_0$, while an order
of magnitude more energy is injected by the stronger quench, with an excess energy of $\Delta E=0.54\,E_0$.
More energy is available on the latter case to excite maxons, which have about twice the energy of rotons,
see Fig.\ref{FIG:3D199eps}.

\subsection{Comparison with tVMC}
\label{ssec:compare}

We compare \thnc\ results for the quench dynamics with corresponding results obtained
with time-dependent variational Monte Carlo (tVMC) simulations~\cite{carleo2012sr}.
Details about our implementation of tVMC can be found in the appendix and in Refs.~\onlinecite{gartner2023, gartnerSciPost22}.
We use the same Jastrow ansatz~(\ref{eq:Phi}) as in the \thnc\ method. However,
tVMC does not require an approximation for $g_3$ nor does tVMC need to
use approximations for the elementary diagrams, because all integrations over the $N$-body configuration
space are performed by brute force Monte Carlo sampling.
The price is, of course, a much higher computational cost. Therefore we have restricted the
comparisons with tVMC to one and two dimensions.

\begin{figure}[ht]
\includegraphics*[width=0.49\textwidth]{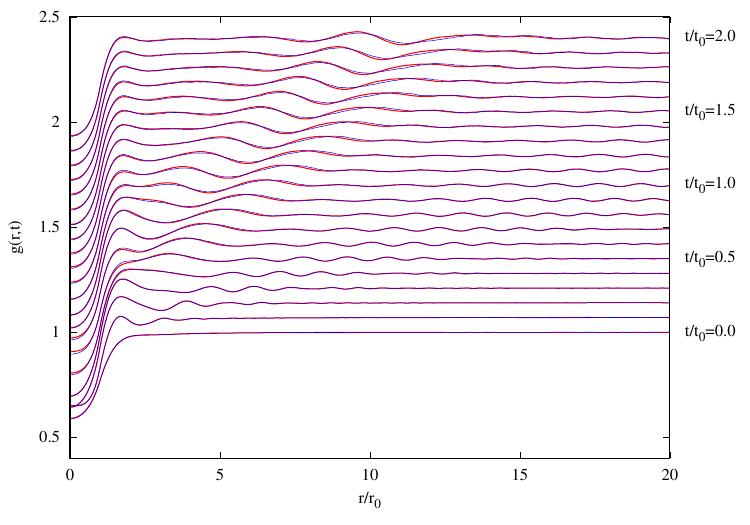}
\includegraphics*[width=0.49\textwidth]{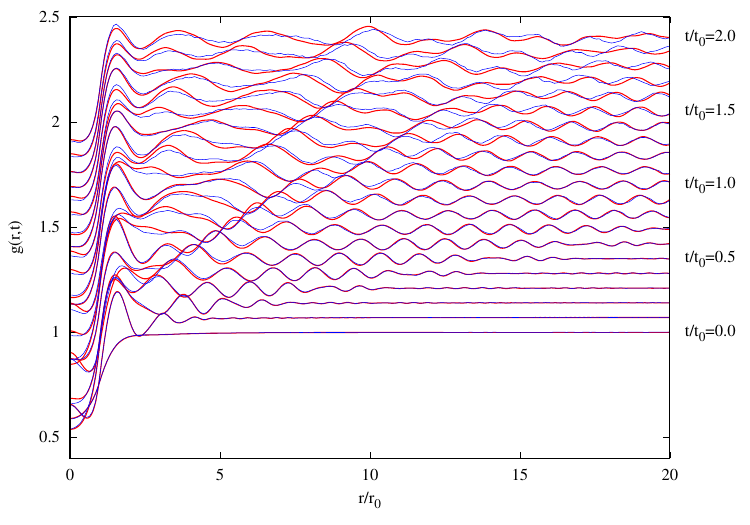}
\caption{
Comparison of the pair distribution function $g(r,t)$ for a Bose gas in 1D between
\thnc\ (red) and tVMC (blue). The interaction is the square well potential~\eqref{eq:vSW}, quenched from
initial strength $U=2\,E_0$ to a strength $U=4\,E_0$ (left panel) and $U=8\,E_0$ (right panel), respectively.
For better visibility $g(r,t)$ is shifted in proportion to time, indicated on the right margin.
}
\label{FIG:1Dcomp}
\end{figure}

\subsubsection{Comparison in 1D}

For the 1D comparison we use a square well interaction potential
\begin{align}
   v(r,t) &= U(t) \Theta\left[R(t)-r\right]
   \label{eq:vSW}
\end{align}
characterized by strength
$U(t)$ and range $R(t)$. The length and energy units are $r_0=\rho^{-1}$ and $E_0={\hbar^2\over 2mr_0^2}$.
In the tVMC simulations we use $N=100$ particles, corresponding to a simulation box size of~$L=100\,r_0$.
We can thus calculate~$g(r,t)$ up to a maximal distance of~$r=50\,r_0$; when fluctuations reach this maximal
distance, spurious reflections appear due to the periodic boundary conditions.
Therefore, we restrict our comparisons of $g(r,t)$ with \thnc\ to times~$t$ before these reflections become noticeable.
Further technical details of the tVMC simulations can be found in the appendix.

In Fig.~\ref{FIG:1Dcomp} we compare $g(r,t)$ after a quench. The interaction range is fixed at $R/r_0=1$
and the interaction strength jumps from $U=2\,E_0$ to a target value $U=4\,E_0$ (left panel) and to $U=8\,E_0$ (right panel).
We show $g(r,t)$ at times $t/t_0=0.0; 0.1; 0.2;\dots;2.0$.
For the weaker first quench, the agreement between \thnc\ (red) and tVMC (blue) is excellent, because the target
interaction is weak enough that neglecting elementary diagrams and the Kirkwood superposition approximation~(\ref{eq:kirkwood})
for the three-body distribution $g_3$ are still good approximations.
For the stronger quench to a target value $U=8\,E_0$ the \thnc\ and tVMC results for $g(r,t)$ do
not match perfectly anymore. For strong interaction, the effect of elementary diagrams or
the three-body distribution or both becomes more important. But overall, the
agreement is still remarkably good; for example, both frequency and phase shift of
the oscillations in $g(r,t)$ are the same. We conclude that \thnc\ works quite well compared to tVMC in 1D,
despite the approximations that we use in our simple implementation of \thnc . Of course, for
strong interactions, even tVMC with pair correlations is not sufficient for quantitative predictions
of the dynamics in 1D and a better variational ansatz than \eqref{eq:Phi} should be used.

\begin{figure}
\begin{center}
\includegraphics*[width=0.68\textwidth]{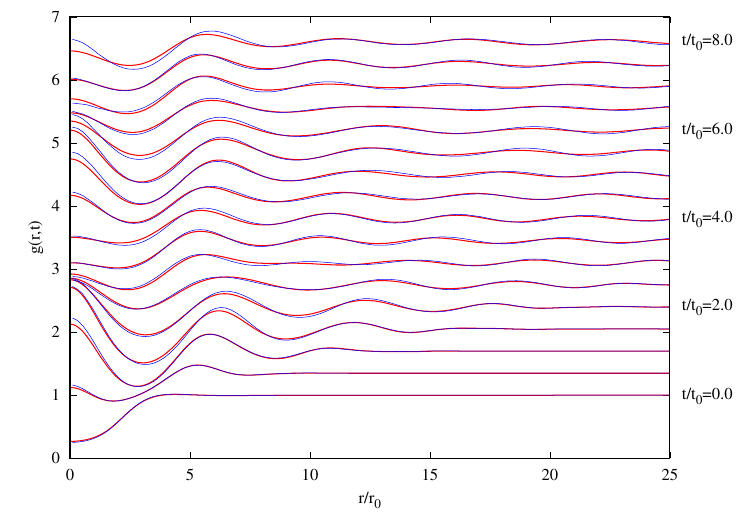}
\end{center}
\caption{
Comparison of the pair distribution function $g(r,t)$ for a Bose gas in 2D between
\thnc\ (red) and tVMC (blue). The interaction is the Rydberg potential~\eqref{eq:vRyd}, quenched from
initial range $R=2\,r_0$ to $R=4\,r_0$.
For better visibility $g(r,t)$ is shifted in proportion to time, indicated on the right margin.
}
\label{FIG:rydberg2Dcomp}
\end{figure}

\subsubsection{Comparison in 2D}

For the 2D comparison we use the Rydberg-dressed interaction~\eqref{eq:vRyd} from the 3D studies in the previous
section~\ref{ssec:3D}. We use the 2D version of the units introduced above for 3D, see also~\cite{seydi_rotons_2020}:
wave numbers are in units of $k_0=(4\pi\rho)^{1/2}$, and again length in units of $r_0=1/k_0$, energy in units
of $E_0={\hbar^2k_0^2\over 2m}$, and time in units of $t_0={\hbar\over E_0}$.
Again, we compare only up to times before
effects of the periodic boundaries in tVMC contaminate the dynamics of $g(r,t)$.

\begin{figure}[ht]
\begin{center}
\includegraphics*[width=0.68\textwidth]{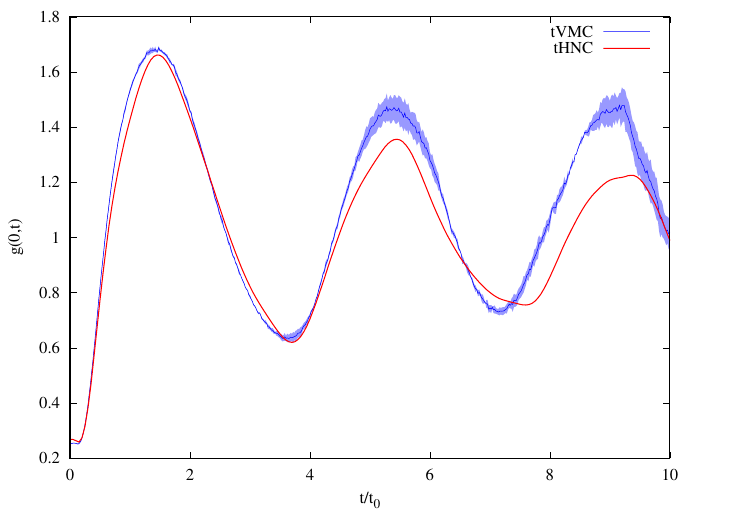}
\end{center}
\caption{
For the quench of the 2D Bose gas in Fig.~\ref{FIG:rydberg2Dcomp}, we compare the roton-induced oscillation of $g(r,t)$
for $r=0$ between \thnc\ (red) and tVMC (blue). The shaded area shows the stochastic error of the tVMC result.
}
\label{FIG:rydberg2Dcomp0}
\end{figure}

Following the quench procedure in our 3D \thnc\ calculations above,
we keep $U$ fixed at $U_0=2\,E_0$, and switch $R(t)$ from $R_0=2\,r_0$, where
excitations from the ground state are monotonous, to $R_1=4\,r_0$, where the excitation spectrum exhibits rotons.
Fig.~\ref{FIG:rydberg2Dcomp} compares the \thnc\ result (red) and tVMC result (blue) for $g(r,t)$ after the quench.
As in 1D, we find good agreement between \thnc\ and tVMC before spurious oscillations due to the periodic boundary
conditions in tVMC appear for later times (not shown). As expected from our 3D calculations, the creation of
pairs of rotons leads to persistent oscillations in $g(r,t)$ for small $r$. The amplitude, however, becomes smaller
in the \thnc\ results, as can be seen in Fig.~\ref{FIG:rydberg2Dcomp0}, which compares $g(r,t)$ for $r=0$. The deviation
between the \thnc\ result and the tVMC result is larger than the stochastic error inherent in the tVMC
method, which increases with time and is shown as shaded area in Fig.~\ref{FIG:rydberg2Dcomp0}.
Hence, the differences between \thnc\ and tVMC are due to the approximations made in our present implementation of \thnc{}. 
When $g(0,t)$ becomes large, particles tend to cluster together, and we expect that
particularly the Kirkwood superposition form~(\ref{eq:kirkwood}) for $g_3$ is a poor approximation.


\section{Discussion}

We present a new method for studying quantum many-body dynamics far from equilibrium,
the time-dependent generalization of the hypernetted-chain Euler-Lagrange method, \thnc ,
which is non-perturbative and goes beyond the mean field paradigm.
We demonstrate this variational method in a study of the interaction
quench dynamics of a homogeneous Rydberg-dressed Bose gas. A sudden
strong change of the effective Rydberg interaction leads to a strong response of the
pair distribution function $g(r,t)$, which is the quantity we are interested in in this work.
In an interaction quench, the Hamiltonian becomes time-dependent but is still translationally
invariant. The systems stay homogeneous and only pair quantities like $g(r,t)$ carry the
dynamics (we assume there is no spontaneous breaking of translation invariance).

We derived the Euler-Lagrange equations of motion for 
$g(r,t)$ for the simplest case, where elementary diagrams in the HNC relations are omitted,
and only pair correlations~$u_2$ are taken into account in the variational ansatz
for the many-body wave function (Jastrow-Feenberg ansatz). This simple version is long
known in various formulations for ground state calculations~\cite{Camp,lanttosiemens} and is sufficient for
not too strongly correlated Bose systems (but not sufficient for quantitative
predictions of energy and structure of e.g. liquid~$^4$He).

The dynamics of the pair distribution $g(r,t)$ of a Rydberg-dressed 3D gas is similar to results of previous tVMC studies
of bosons on a deep 2D lattice, described by the Bose-Hubbard model~\cite{carleo2014pra}, and
bosons in 1D~\cite{carleo2017prx}. After a weak quench a ``wave'' travels to larger distances. Unlike for the
dynamics on a lattice, we observe no light-cone restriction for the speed of propagation since our
group velocity is not bounded from above. Ripples of very high wave number indeed move very
fast towards larger distances. We stress that these waves are not density fluctuations (which would
break translation symmetry) but fluctuations of the {\em pair} density.

There is a qualitative change in the behavior of $g(r,t)$ for a strong quench.
If the interaction after the quench is strong enough that a linear response calculation predicts
a roton excitation, i.e. a non-monotonous dispersion relation, $g(r,t)$ exhibits long-lasting
oscillations for small $r$ that do not decay within our calculation time windows. This can
be readily understood as the creation of a pair of rotons with opposite momentum (total momentum
is conserved by an interaction quench). The group velocity of rotons vanishes, therefore the oscillations
of $g(r,t)$ due to rotons do not propagate to larger $r$; furthermore the density of states
for rotons diverges, therefore exciting roton pairs is very efficient. The argument is equally valid
for the maxon, i.e. the local maximum of the dispersion relation. Finally, since the quench creates
a pair of rotons or maxons (a roton-maxon pair would violate momentum conservation), the
oscillations have twice the roton or maxon frequencies. Indeed, for these strong quenches the power spectra
have peaks at almost these frequencies, and not much strength at other frequencies. The peaks are
slightly off from twice the roton or maxon frequency due to nonlinear effects, which we confirmed
by smaller, hence more linear-response-like, jumps in the interaction quench.

Apart from the Jastrow-Feenberg ansatz with pair correlations, the \thnc\ equations contain
two approximations, the already mentioned omission of elementary diagrams, and an
approximation for the three-body distribution function $g_3(\qr_1,\qr_2,\qr_3)$, where we
chose the straightforward Kirkwood superposition approximation. In order to assess these
two approximations, we performed tVMC simulations of interaction quenches. We restricted
ourselves to one and two dimensions due to the high computational cost of tVMC.
Overall we see good agreement between \thnc\ and tVMC. In 1D we compared for two different quenches,
finding excellent agreement for the weaker quench. As the interaction strength after the
quench increases the agreement worsens somewhat, but all the main features such as the
roton-induced oscillations in the 2D comparison are captured already by \thnc .

Comparison with tVMC or other time-dependent many-body methods may be prohibitive, such as for
the study of the long-range behavior of $g(r,t)$ in 3D. In such a case, the errors in \thnc\
can only be studied and reduced by improving \thnc\ itself.
For strong interactions, we should (i) incorporate elementary diagrams -- this can be done
approximately in the same way as in ground state HNC-EL calculations;
(ii) improve upon the Kirkwood superposition approximation (e.g.\ use the systematic Abe expansion~\cite{feenberg1969})
or compare with other approximations like the convolution approximation~\cite{HaM76}; (iii) include
triplet correlations $u_3(\qr_1,\qr_2,\qr_3)$ in the variational ansatz (also for tVMC), as has
been done for ground state and linear response calculations~\cite{campbell2015prb}.
The latter improvement of \thnc\ will require completely new approaches how to solve the resulting
high-dimensional equations of motion.

Like the ground state HNC-EL method, \thnc\ can be generalized to inhomogeneous and/or
anisotropic systems. The latter is important for dipolar Bose gases, while most quantum gas
experiments are in harmonic or optical traps rather than in box traps, and thus require an
inhomogeneous description; even in case of an interaction quench of a homogeneous Bose
gas, the quench may trigger spontaneous breaking of translational invariance.
The diagrammatic summations used in the ground state HNC-EL method have been
generalized to off-diagonal properties like the one-body density matrix to study off-diagonal
long-range order, i.e.\ the Bose-Einstein condensed fraction~\cite{manousakis_condensate_1985}. We will generalize
this to the time-dependent case.
Further down the road, we plan to include three-body correlations $u_3$ at least approximately, known to
improve ground states of highly correlated systems~\cite{Kro86}. This
opens a new scattering channel, where an interaction quench can generate excitation
triplets, not just pairs, with total momentum zero. In the case of quenches to rotons,
the oscillation pattern of $g(r,t)$ will be more complex because its power spectrum can
contain frequencies corresponding to the energies of three rotons.
We are also generalizing \thnc\ to anisotropic interaction in order to study
the non-equilibrium dynamics of dipolar quantum gases.
Studying the long-time dynamics after a nonlinear perturbation, such as a strong interaction
quench, requires numerical stability for very long times, which is challenging for
nonlinear problems like solving the \thnc\ equations or the equations of motion of
tVMC. A small time step is required
to achieve long-time stability for both \thnc\ and tVMC. However, \thnc\ is
computationally very cheap and calculations of the long-time dynamics are feasible.
In order to delay artifacts from reflections at domain boundaries,
either the spatial domain must be chosen very large, or absorbing boundary conditions are
implemented.


\section*{Acknowledgements}
We acknowledge discussions with Eckhard Krotscheck, Markus Holzmann, and Lorenzo Cevolani.
The tVMC simulations were supported by the computational resources of the Scientific Computing
Administration at Johannes Kepler University.


\begin{appendix}

\section{Derivation of the \thnc\ equation}

We derive the \thnc\ equation of motion~(\ref{eq:tHNCel}) using the time-dependent
variational principle $\delta \mathcal{S}=0$, where $\mathcal{S}$ is the action defined in eq.(\ref{eq:defS}).

The action is given by eqns.~(\ref{eq:defS}), (\ref{eq:defL}), and~(\ref{eq:lagrangian}) as
\begin{align*}
    \mathcal{S} = \mathcal{S}_g + \mathcal{S}_2 + \mathcal{S}_3
\end{align*}
with
\begin{align*}
    \mathcal{S}_g &= {\rho\over 2} \int\limits_{t_0}^{t}\!dt'\!\! \int\!\! d^dr \Big[ v(r,t')
  - {\hbar^2\over 4m} \Delta u(r,t') \Big] g(r,t'),\\
    \mathcal{S}_2 &=  {\hbar\over 2} \int\limits_{t_0}^{t}\!dt'\!\! \int\!\! d^dr\, \dot{\varphi}(r,t')\, g(r,t') 
  +{\hbar^2\over 2m} \rho\! \int\limits_{t_0}^{t}\!dt'\!\! \int\!\! d^dr\, \qv(\qr,t')^2 g(r,t'), \\
    \mathcal{S}_3 &=   {\hbar^2\over 2m} \rho^2\! \int\limits_{t_0}^{t}\!dt'\!\! \int\!\! d^drd^dr'\,
    \qv(\qr,t') \cdot\qv(\qr',t')\,  g_3(r,r',|\qr-\qr'|,t'),
\end{align*}
where ${\qv(\qr,t')\equiv \nabla\varphi(r,t')}$.

The Euler-Lagrange equations~(\ref{eq:deltaSdeltag}) and~(\ref{eq:deltaSdeltaphi}) require
the functional differentiation of $\mathcal{S}_g$, $\mathcal{S}_2$, and $\mathcal{S}_3$ with respect
to $g(r,t)$ and $\varphi(r,t)$. Apart from time integration, $\mathcal{S}_g$ is the same expression
as the energy expectation value of the ground state HNC-EL method. It does not depend on
$\varphi(r,t)$ and the variation with respect to $g(r,t)$, using the HNC relation~(\ref{eq:HNC}),
can be found in reviews on HNC-EL, e.g. Ref.~\onlinecite{QMBT00Polls}. The derivatives of $\mathcal{S}_2$ are
\begin{align*}
{\delta\mathcal{S}_2\over\delta g(r,t)} &= {\hbar\over 2}\dot\varphi(r,t) + {\hbar^2\over 2m}\qv(\qr,t)^2, \\
{\delta\mathcal{S}_2\over\delta \varphi(r,t)} &= -{\hbar\over 2}\dot g(r,t) - {\hbar^2\over 2m}\nabla\left[ g(r,t)\cdot\qv(\qr,t)\right].
\end{align*}

In $\mathcal{S}_3$ we employ the Kirkwood superposition approximation~(\ref{eq:kirkwood})
\begin{align*}
    \mathcal{S}_3 &=   {\hbar^2\over 2m} \rho^2\! \int\limits_{t_0}^{t}\!dt'\!\! \int\!\! d^drd^dr'\,
    \qf(\qr,t') \cdot\qf(\qr',t')\,  g(|\qr-\qr'|,t')
\end{align*}
with $\qf(\qr,t')\equiv g(r,t')\,\qv(\qr,t')$. The derivatives are
\begin{align*}
{\delta\mathcal{S}_3\over\delta g(r,t)} &= {\hbar^2\over m}\qv(\qr,t)\cdot\nabla\gamma(r,t) + {1\over 2} \beta(r,t), \\
{\delta\mathcal{S}_3\over\delta \varphi(r,t)} &= -{\hbar^2\over m}\nabla\left[g(r,t)\cdot\nabla\gamma(r,t)\right].
\end{align*}
where $\beta(r,t)$ and $\nabla\gamma(r,t)$ are defined in eqns.~(\ref{eq:beta}) and~(\ref{eq:gamma}),
respectively.

Putting everything together, equation~(\ref{eq:deltaSdeltag}) becomes, after multiplying by 2,
\begin{align}
  0 =&- {1\over\sqrt{g(r,t)}}{\hbar^2\over m}\nabla^2\sqrt{g(r,t)}  + v(r,t) + w_I(r,t) \nonumber\\
  &+ \hbar\dot\varphi(r,t) + {\hbar^2\over m}[\nabla\varphi(r,t)]^2  \nonumber\\
  &+ 2{\hbar^2\over m} \nabla\varphi(r,t)\cdot\nabla\gamma(r,t) + \beta(r,t).
  \label{eq:deltaSdeltag2}
\end{align}
If we kept only the first line of the equation, we would recover the ground state HNC-EL equation (where
$\varphi=0$, of course). The induced potential $w_I(r,t)$ describing phonon-mediated interactions can
be found in Ref.~\onlinecite{QMBT00Polls}.
Equation~(\ref{eq:deltaSdeltaphi}) becomes
\begin{align}
  0 &= -{1\over 2}\hbar\dot g(r,t) - {\hbar^2\over m}\nabla\left[ g(r,t)\cdot\nabla(\varphi(r,t) + \gamma(r,t)\right].
  \label{eq:deltaSdeltaphi2}
\end{align}
When we multiply eq.~(\ref{eq:deltaSdeltag2}) with $\sqrt{g(r,t)}$ and eq.~(\ref{eq:deltaSdeltaphi2}) with
${i / \sqrt{g(r,t)}}$, add the two equations, and multiply the resulting equation with $e^{i\varphi(r,t)}$,
we obtain the final form~(\ref{eq:tHNCel}) of the \thnc\ equation for the effective
pair wave function $\psi(r,t)=\sqrt{g(r,t)}e^{i\varphi(r,t)}$ that we solve numerically.

\section{The tVMC method}
\label{sec:tVMC}

In tVMC~\cite{carleo2012sr,carleo2014pra,carleo2017prx}, we use the same Jastrow-Feenberg ansatz of equation \eqref{eq:Phi} as in tHNC, and describe the time dependence of the wavefunction
via a set of $P$ complex variational parameters ${\alpha(t) = \{\alpha_1(t), \alpha_2(t), \ldots, \alpha_P(t) \}}$,
which are coupled to local operators ${\mathcal{O}_m(\qr_1, \ldots, \qr_N)}$. In our implementation
(more details can be found in~\cite{gartnerSciPost22}), these local operators are real and represented by third
order B-splines $B_m(r)$ centered on a uniform grid in the interval $[0,L/2]$, and are given by
${\mathcal{O}_m(\qr_1, \ldots, \qr_N) = \sum_{k<l} B_m(r_{kl})}$.
The real and imaginary part of the pair correlation function $u_2(r)$ can then be written as
\begin{align*}
    u(r) &=\sum_m^P B_m(r) \alpha_m^R(t) &
    \varphi(r) &= \sum_m^P B_m(r) \alpha_m^I(t) ,
\end{align*}
where $\alpha_m^R$ and $\alpha_m^I$ are the real and imaginary part of~$\alpha_m$. The time evolution of
the variational wavefunction is obtained by solving the coupled system of equations (see~\cite{carleo2012sr})
\begin{align}
    i \sum_{n} S_{mn} \dot \alpha_n = \langle \mathcal{E}\mathcal{O}_m \rangle - \langle \mathcal{E} \rangle \langle \mathcal{O}_m \rangle ,
    \label{eq:tVMCEOM}
\end{align}
with the correlation matrix ${S_{mn} = \langle \mathcal{O}_m \mathcal{O}_n \rangle - \langle \mathcal{O}_m \rangle \langle \mathcal{O}_n \rangle}$ and the local energy ${\mathcal{E}=\frac{H |\Phi\rangle}{|\Phi\rangle}}$.
The expectation values $\langle \ldots \rangle$ are estimated using Monte Carlo integration by sampling
from the trial wavefunction ${\Phi(\qr_1, \ldots, \qr_N, t) = \exp \left( \sum_{k<l} u_2(r_{kl},t) \right)}$.
Because of the translational invariance of the studied system we do not need to take into account
a one-body part $u_1(r, t)$ in the wavefunction, unlike in Ref.~\cite{gartnerSciPost22} where the
dynamics of a Bose gas in a 1D optical lattice has been simulated.

\medskip{}
\noindent
\textbf{Simulation parameters in 1D: }
In tVMC, we approximate the Jastrow pair-correlation function~$u(r)$
using a cubic spline function with $P=400$ complex weights,
corresponding to the time-dependent variational parameters $\alpha_m(t)$, as given above.
The time propagation is performed with a time step of $\Delta t=2\cdot10^{-4}\,t_0$, and ${N_{\text{MC}} = 1250}$
uncorrelated samples are used for calculating the expectation values required for time propagation. The results are
converged with respect to the spatial and temporal numerical resolution as we see no changes in the simulation results upon
increasing $P$ or decreasing $\Delta t$.

Performing tVMC simulations comes of course with a large computational workload compared to tHNC calculations. While about 560 CPU hours are required to run the discussed tVMC simulations in 1D, tHNC takes about 5 CPU minutes for evolving the pair distribution function for the
same time window $t\in[0,20]t_0$ and for an even larger $r$-domain, which reduces unphysical boundary reflections.
However, as most Monte Carlo methods, tVMC can be highly parallelized, which makes it feasible in terms of real computing time.

\medskip{}
\noindent
\textbf{Simulation parameters in 2D: }
In the 2D tVMC simulations we use $N=900$ particles, corresponding to a simulation box size $L=106.347\,r_0$.
We use ${P=300}$ variational parameters, a time step of ${\Delta t = 2\cdot10^{-4}\,t_0}$
and ${N_{\text{MC}} = 1250}$ uncorrelated samples for estimating the expectation values needed to solve equation (\ref{eq:tVMCEOM}). 

For the 2D tVMC simulations with $N=900$ particles the computational time amounts to 14.000 CPU hours, stressing one more time the benefits of tHNC simulations concerning the computational workload, which only amounts to approx.\ 30 CPU hours.

\section{Comparison with Bogoliubov approximation}
\label{sec:bogo}

Interaction quenches in Bose gases have been studied with the Bogoliubov method which
accounts for correlations as a perturbation (``quantum fluctuations'') of the mean field approximation.
The pair distribution function $g(r)$ in Bogoliubov approximation~\cite{lee_eigenvalues_1957} has
been generalized to dynamical problems in Refs.~\cite{natu_dynamics_2013,natu_dynamics_2014}. Particularly for
quenches in Rydberg-dressed Bose gases this has been used, with a further approximation,
in Ref.~\cite{mccormack_dynamical_2020}.

\begin{figure}[ht]
\begin{center}
\includegraphics*[width=0.68\textwidth]{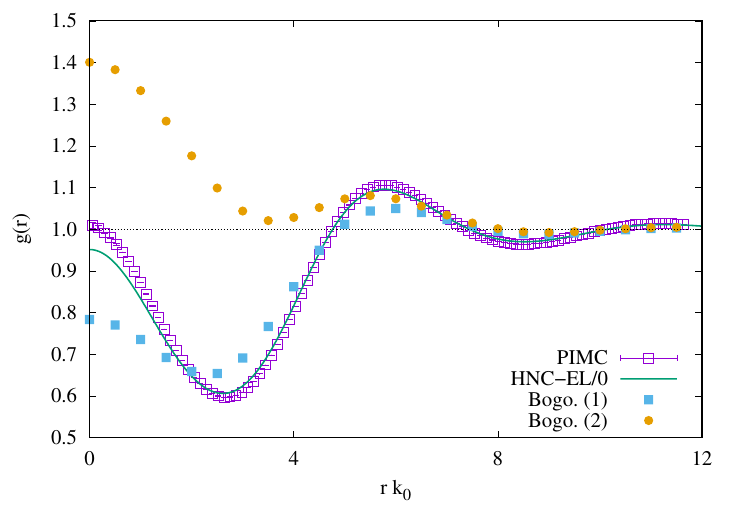}
\end{center}
\caption{
Comparison of the ground state pair distribution function $g(r)$ obtained from HNC-EL (line),
from exact PIMC in the limit of low temperature (open squares), from the Bogoliubov method
(filled circles), and an approximate Bogoliubov method (filled squares). See text for details.
}
\label{FIG:gbogo}
\end{figure}

In Fig.~\ref{FIG:gbogo}
we compare the result for the ground state $g(r)$ for the 3D Rydberg-dressed Bose gas for $R=4$ and $U=3$ obtained with
four different methods: exact path integral Monte Carlo (PIMC) in the limit of low temperature, such
that the Bose gas is effectively in the ground state~\cite{seydi_rotons_2020}; the HNC-EL/0 method which is the
ground state limit of the time-dependent \thnc\ method used in this paper in; and two variants of the
Bogoliubov method: first, the full expression for $g(r)$ used in Refs.~\cite{natu_dynamics_2013,natu_dynamics_2014}
for dynamics, referred to as Bogo.(1) in Fig.~\ref{FIG:gbogo}; and secondly, an approximate expression used
in Ref.~\cite{mccormack_dynamical_2020} for dynamics, referred to as Bogo.(2) in Fig.~\ref{FIG:gbogo}.
From Fig.~\ref{FIG:gbogo} we see that HNC-EL/0 result for $g(r)$ (line) reproduces the exact PIMC result (open squares)
very well, apart from small deviations for $r\to 0$. On the other hand, both Bogoliubov results deviate quite
significantly from the exact PIMC result.
The comparison demonstrates that correlations must be incorporated non-perturbatively in this case, while
for weakly interacting Bose systems the Bogoliubov approximation may be sufficient.

\end{appendix}

\bibliography{bec,papers,literature,zotero,nonequil}

\end{document}